\pdfoutput=1
\documentclass[a4paper,noarxiv]{quantumarticle}

\usepackage[T1]{fontenc}
\usepackage{amsmath,amssymb,amsfonts}
\usepackage{graphicx}
\usepackage{booktabs}
\usepackage{microtype}
\usepackage{xcolor}
\usepackage{hyperref}

\graphicspath{{./}}

\begin{document}

\title{The verifier side of speculative window decoding:
a predictability bracket, a machine-checked blast-radius bound,
and a decoder-agnostic recover loop}

\author{Rylan Malarchick}
\affiliation{Department of Engineering Physics,
             Embry-Riddle Aeronautical University, Daytona Beach, FL 32114, USA}
\email{malarchr@my.erau.edu}

\date{}

\begin{abstract}
Speculative window decoders hide quantum error-correction decoder latency by guessing
the cross-boundary decisions that link adjacent decoding windows, running downstream
work on the guess, and verifying lazily. SWIPER and ARTERY each build one predictor,
about $90\%$ accurate; neither built the verifier side. We build it on a reconstructed
SWIPER harness (Stim rotated surface code, minimum-weight matching). A predictor-only
bracket shows the cross-boundary decision is local, the achievable accuracy reaching
about $0.999$ within three rounds, with small, diffuse headroom over SWIPER. We
establish a worst-case temporal blast-radius bound, its probability core
machine-checked in Lean~4 and conditional on a modeling reduction we then test: a
misprediction's effect decays exponentially in the commit width, so the radius is one
and speculation adds no error floor. We falsify that reduction shot by shot and find
the real mechanism, clearest at near-threshold noise, is a global minimum-weight
re-pairing. A compiler pass derives SWIPER's restart policy from these numbers; a
runtime executor confirms on the harness that the loop recovers exactly and removes the
serial commit-chain stall up to a small penalty. A second decoder (union-find) settles
which results are decoder-agnostic: the predict-verify-recover wrapper and the
structural phenomenology, while the absolute magnitudes and the min-weight mechanism
are matching-specific.
\end{abstract}

\maketitle

\section{Introduction}

Real-time quantum error correction has a latency problem. Syndrome rounds arrive on a
fixed hardware cadence, about a microsecond per round for superconducting
qubits~\cite{caune_realtime,google_belowthreshold}, and a decoder that cannot keep pace
lets the syndrome backlog grow until the race against decoherence is
lost~\cite{battistel_realtime,terhal_review}. Window decoding splits the long syndrome
history into windows that can be decoded in
parallel~\cite{parallelwindow,topological_memory}, but adjacent windows share a
boundary: the correction committed in one window determines the decode problem in the
next. That dependency is a serial bottleneck.

Speculation removes it. SWIPER~\cite{swiper} predicts the cross-boundary decode
decision, lets the downstream window start on the prediction, and runs the full
decoder lazily to verify; a wrong prediction ``poisons'' the window and triggers a
restart. ARTERY~\cite{artery} applies the same idea to feedback gates with a literal
branch predictor. This is the structural twin of CPU branch prediction (predict,
execute, flush) and of large-language-model speculative
decoding~\cite{leviathan,spec_sampling} (draft, verify, roll back). Both quantum
systems report the empirical accuracy of one hand-built predictor, about $90\%$, and a
runtime improvement of about $40\%$ for SWIPER~\cite{swiper,artery}.

What neither system built is the verifier side. Predicting is the generator half of
speculation; verifying is the other half, and it raises four questions that nobody has
answered. How good could a guess be, and is $90\%$ near a limit or leaving points on
the floor? How far can a wrong guess corrupt the computation before the lazy verify
catches it, in the worst case where the guess is wrong precisely where the boundary was
least predictable? Given those two numbers, is speculation worth it at a given
boundary? And does executing the full predict, verify, and recover loop on a real
decoder actually hide latency and recover correctly? These sit above the decoder, so
they do not compete with work on faster decoders or
predecoders~\cite{promatch,pinball}; the verifier side is a reusable layer that any
decoder plugs into behind the verify step.

We answer the four on a reconstructed SWIPER harness. We measure the predictability
(Sec.~\ref{sec:headA}) and find the decision local and SWIPER near the achievable
accuracy; we bound the blast radius with a machine-checked probability core, then
falsify and replace its modeling hypothesis (Sec.~\ref{sec:headB}); we feed both
numbers to a compiler pass that derives SWIPER's restart policy
(Sec.~\ref{sec:headC}); we run the whole loop on the harness and measure recovery and
latency-hiding (Sec.~\ref{sec:headD}); and we re-run the decoder-dependent findings
under a second decoder to state precisely which results are decoder-agnostic
(Sec.~\ref{sec:relativity}).

\section{Setup and the calibration gate}
\label{sec:setup}

The exact SWIPER-SIM generator configuration is not public, so the harness is a
documented reconstruction pinned in code. We generate rotated surface-code memory
circuits in Stim~\cite{stim} (\texttt{surface\_code:rotated\_memory\_z}) with uniform
circuit-level depolarizing noise at rate $p$ on all four channels. A window spans $2d$
rounds, with a commit region of the first $d$ rounds and a buffer of the last $d$; the
boundary cut is the detector layer at $t=d$. The reference decoder is minimum-weight
perfect matching (MWPM) via PyMatching~\cite{pymatching}. The boundary (dependency)
bits, the quantity a real-time predictor must guess, are the node parities of the
full-decode matched edges that cross the cut, which are exactly the boundary syndrome
bits passed to the next window. Speculation accuracy is the fraction of boundaries
with every dependency bit correct, the metric SWIPER reports.

\begin{figure}[t]
\includegraphics[width=\columnwidth]{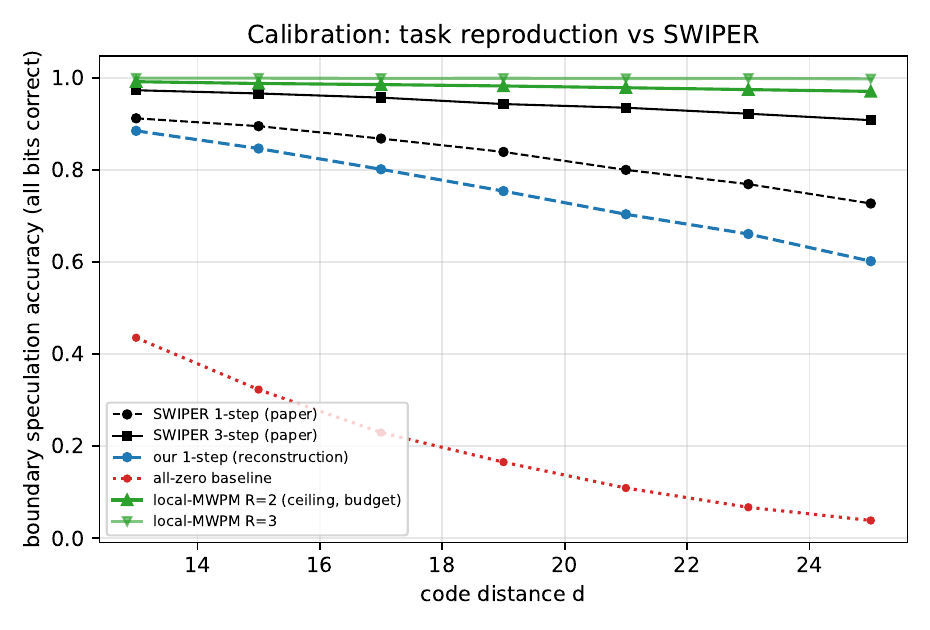}
\caption{Calibration gate. The reconstructed harness reproduces SWIPER's accuracy
regime on the identical boundary-prediction task: radius-2 local MWPM brackets the
reported three-step predictor, and the one-step curve reproduces SWIPER's one-step in
shape.}
\label{fig:calibration}
\end{figure}

A ceiling compared against an accuracy measured on a different task is meaningless, so
the first step is a calibration gate: reproduce SWIPER's accuracy regime on the
identical task before any headroom claim. A radius-2 local MWPM predictor on this
harness reaches $0.992$ to $0.971$ over $d=13$ to $25$, bracketing SWIPER's reported
three-step predictor ($0.973$ to $0.908$); our one-step predictor reproduces SWIPER's
one-step curve in shape, offset a few points low because their exact noise and graph
configuration differs (Fig.~\ref{fig:calibration}). The gate passes. All headroom in
this paper is measured as predictor-versus-bracket on the same harness, so it is
internally consistent regardless of the few-point absolute offset from the paper.

\section{Measure: the predictability and speedup bracket}
\label{sec:headA}

We bracket the best boundary-prediction accuracy achievable from the information a
real-time predictor sees, without a plug-in entropy estimator, which would die to the
curse of dimensionality on a high-dimensional syndrome vector. A predictor's held-out
accuracy is a rigorous lower bound on the Bayes accuracy, and its held-out
cross-entropy is a rigorous upper bound on the conditional entropy $H(Y\mid X)$ by
Gibbs' inequality, giving an entropy-implied accuracy estimate through Fano. The
achievable side is a principled radius-$R$ local MWPM decoder, which uses only the
syndrome within $R$ rounds of the cut and matches chains leaving that band to an
artificial boundary, exactly as a window decoder treats its open boundary. A learned
per-bit gradient-boosted model on the same local stencil corroborates it. We use
``ceiling'' informally for the bracket's upper end; the rigorous content is the
bracket, not a proof over all predictors. A general predictability ceiling, one that
holds over all predictors rather than bracketing this decoder's boundary decisions on
this reconstructed harness, is left to future work. The bracket is validated on a
synthetic
source with closed-form entropy before any real-data claim: it sandwiches the known
Bayes accuracy ($0.712$ true, $0.710$ recovered) and entropy ($1.507$ bits true,
$1.528$ recovered).

\begin{figure}[t]
\includegraphics[width=\columnwidth]{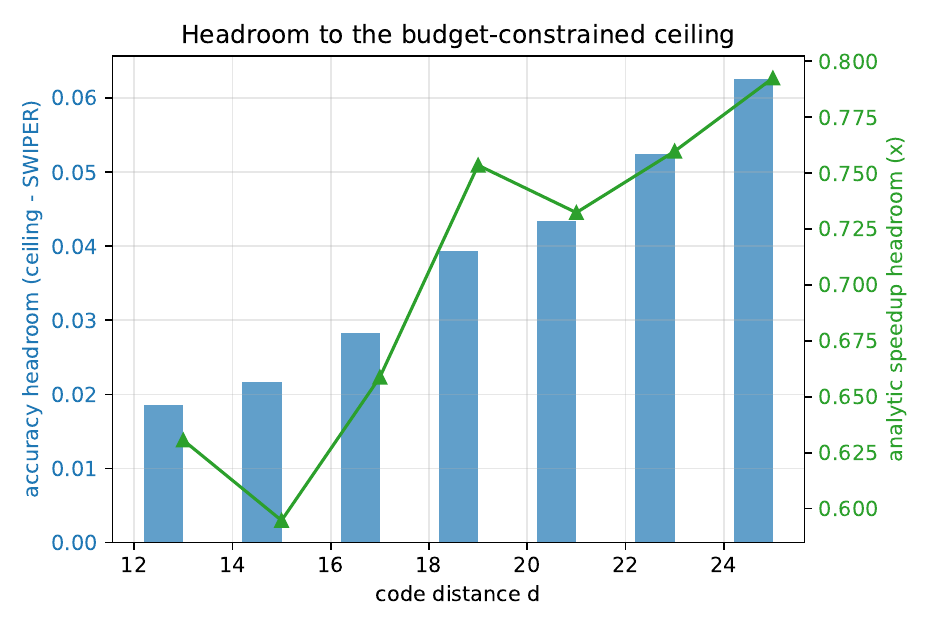}
\caption{The boundary decision is local and the headroom over SWIPER is small but
grows with distance. Achievable (local-MWPM) accuracy saturates by a receptive field
of three rounds; the gap to SWIPER's reported predictor widens from $0.019$ at $d=13$
to $0.063$ at $d=25$.}
\label{fig:headroom}
\end{figure}

The boundary decision is local. The achievable accuracy given a receptive field of $R$
rounds each side of the cut rises steeply and saturates (Fig.~\ref{fig:headroom}):
$R=2$ reaches $0.97$ to $0.99$ across $d=13$ to $25$, $R=3$ reaches about $0.999$, and
$R=4$ is indistinguishable from the full decode. The cross-boundary decision is set by
roughly three rounds of syndrome on each side, so the dominant accuracy lever is the
receptive field, not predictor cleverness.

The headroom over SWIPER is real but modest, and it grows with distance. At SWIPER's
own receptive field the achievable accuracy exceeds the reported three-step predictor
by $0.019$ at $d=13$, rising to $0.063$ at $d=25$.
Mapped to a speedup through the speculative-decoding formula of
Leviathan~\textit{et al.}~\cite{leviathan}, with acceptance rate $\alpha$ equal to the
boundary accuracy and cost ratio $c$ equal to predictor over decoder cost, that is a
relative improvement of $1.19\times$ at $d=13$ to $1.33\times$ at $d=25$, reported as a
band over $c\in[0.05,0.5]$. This map assumes linear-chain rollback and a free
speculation depth $\gamma\le 32$, so the ratios are optimistic in absolute terms; the
robust quantity is the ratio through the identical map, and the gap to the true
graph-structured rollback is exactly the blast radius of Sec.~\ref{sec:headB}.

That headroom is diffuse, not concentrated. The per-boundary-node error of the budget
predictor is small (mean $5\times10^{-5}$ at $d=21$) and uncorrelated with the
pre-registered structural feature, distance to the spatial code boundary (correlation
$+0.04$). The exact-match gap is the compounding of
a per-bit gap of order $10^{-4}$ over hundreds of bits, not a characterizable hard
subset of boundaries. A smarter cheap predictor buys a few diffuse points; the
actionable accuracy lever is one more round of receptive field. Per-bit predictability
is near-saturated, which points the contribution at the recovery cost, the next
section.

\section{Bound, refute, and explain: the misprediction blast radius}
\label{sec:headB}

\subsection{The temporal containment bound}

A misprediction passes wrong dependency bits into a window. The blast-radius question
is whether that corruption propagates: does the wrong incoming boundary change the
window's outgoing boundary, and so poison the next window? We model exactly that
operator on a single window with open temporal ends: perturb the incoming (sink)
boundary, re-decode with MWPM, and read the change in the outgoing dependency bits at
the cut, $W$ rounds away.

We first fix the adversary class, before any bound, so that the result is neither
vacuous nor trivial. The class $\mathcal{A}(K,\Lambda)$ flips at most $K$ sink bits
within a spacetime ball of diameter $\Lambda$, placed worst-case (wrong precisely where
the boundary was least predictable). Calibration from Sec.~\ref{sec:headA} sets the
parameters rather than assuming them: only $0.3\%$ to $0.6\%$ of boundaries are
mispredicted, a poisoned boundary carries on average $1.04$ wrong bits, and the
maximum over $d=7$ to $11$ is $K=3$. We bound this class above by \texttt{flip\_all},
every sink bit flipped, which is verified to dominate the localized budget-$K$ clusters
empirically, so containment for \texttt{flip\_all} implies containment for the
hard-syndrome-correlated adversary.

\begin{figure}[t]
\includegraphics[width=\columnwidth]{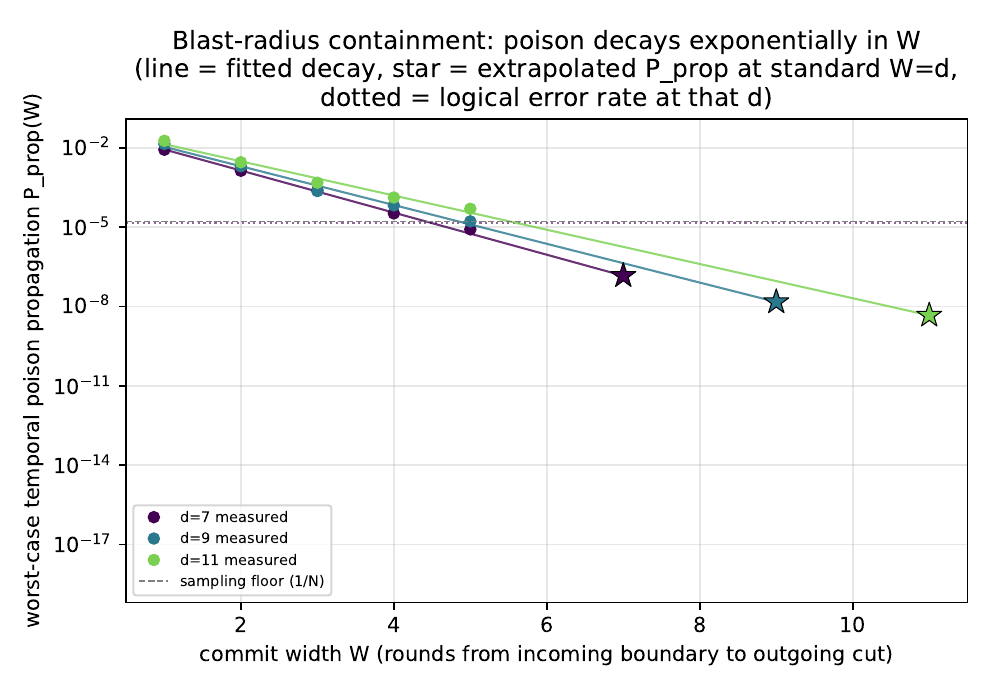}
\caption{Worst-case temporal poison propagation $P_{\mathrm{prop}}(W)$ decays
exponentially in the commit width $W$, with rate $\lambda<1$ at every distance and a
$d^2$ prefactor. At $W=d$ the failure sits well below the logical error rate, so the
blast radius is one.}
\label{fig:blast}
\end{figure}

The finding is that the corruption is contained to the poisoned window. The
propagation probability decays exponentially in the commit width $W$,
\begin{equation}
P_{\mathrm{prop}}(W)\ \le\ C\, d^2\, \lambda^{W},
\qquad \lambda = \sqrt{\deg^2\, c\, p} < 1,
\label{eq:bound}
\end{equation}
with $\deg=6$ the degree of an interior node in the rotated surface-code matching graph
(four spatial and two temporal neighbors~\cite{fowler_surface}) and $c$ an order-one
counting constant. The measured rate is $\lambda = 0.16, 0.19, 0.23$ at $d=7,9,11$
($p=10^{-3}$), each below the conservative rigorous rate $0.38<1$, and the prefactor
scales as $d^2$ (Fig.~\ref{fig:blast}). At the standard commit width $W=d$ the failure
probability is $1.5\times10^{-7}$, $1.5\times10^{-8}$, $4.7\times10^{-9}$ at $d=7,9,11$.
For comparison, the $d$-round logical error rate measured on the same harness is
$1.5\times10^{-5}$ at $d=7$ (below the resolution of our shot count for $d\ge 9$), so at
$d=7$ the propagation sits two orders of magnitude below it, and $P_{\mathrm{prop}}(W=d)$
falls faster in $d$ (as $\lambda^{d}$) than the logical error rate. The temporal blast
radius is one, and verified speculation adds no error floor.

\subsection{Machine-checked core, certified reduction}

We keep the line between proven and empirical explicit. The probability machinery is
machine-checked in Lean~4 with Mathlib~\cite{mathlib}: the development builds with no
\texttt{sorry}, and \texttt{\#print axioms} on the assembled theorem reports only the
three standard axioms (propositional extensionality, choice, quotient soundness). What
Lean proves is the Peierls/union-bound tail (the probability that some boundary-to-cut
path is entirely faulty is at most the path count times $q^m$), the assembled
containment statement (under a reduction and a path-count bound, the propagation
probability is at most $B\,q^{W}$), the subsidized $q^{W/2}$ rate, and the path-count
bound itself (at most $\deg^{W}$, proved with an automated prover). This is the
machinery that makes the bound exponential, with the rate and the constant pinned.

What is not proved from the decoder, and is carried as an explicit hypothesis backed by
the empirical certificate of Eq.~\eqref{eq:bound}, is the reduction itself: that poison
reaching the cut implies a faulty boundary-to-cut path of length at least $W$, together
with the surface-code constants. The clean deterministic version of that reduction (a
weight-$K$ perturbation confines its effect within graph distance $K$) is false under
noise, since a single flipped sink bit changes the outgoing boundary at distance two or
three by riding background error chains, which is why the statement must be
probabilistic. That probabilistic form is exactly what the Lean tail bound consumes.

\subsection{Both halves of SWIPER's observation}

SWIPER reports no degradation between non-adjacent boundaries and about $4\%$ between
adjacent ones. The temporal result above is the first half (non-adjacent, about
$10^{-7}$). For the second half we build a corner window with a spatial cut and a
temporal cut meeting at a corner, perturb the spatial face, and measure the temporal
face. The coupling is a bounded constant of a few percent ($4.96\%$, $5.36\%$, $6.42\%$
at $d=9,11,13$, against SWIPER's $4\%$), invariant to the commit width, and
corner-localized: a single spatial row perturbed at distance $\delta$ from the corner
couples as roughly $\exp(-c\,\delta)$. The two faces meet at an order-one distance, so
the coupling is a constant rather than the $\exp(-cW)$ of the temporal direction. This
reproduces and explains why adjacent is about $4\%$ and non-adjacent is about zero, and
justifies the ``restart poisoned plus adjacent'' policy. We characterize this coupling
but do not prove a containment bound for it, unlike the temporal direction
(Sec.~\ref{sec:limits}).

\subsection{Refuting the reduction, and the real mechanism}
\label{sec:falsify}

The Lean core is conditional on the reduction hypothesis, certified only in aggregate.
We test it shot by shot. From the same decomposed detector error model the decoder
matches against, we extract each shot's ground-truth faults, anchored by an exact
identity: the parity of a shot's fired-fault detector components equals Stim's sampled
syndrome, so the fault set is ground truth, not an inference. A pre-registered battery
of candidate necessary-structure predicates, the two literal forms of the Lean
hypothesis plus variants from three independent lenses (percolation, matching cost,
graph cut), is then searched for counterexamples: a shot that propagates while the
predicate is false refutes that predicate as a necessary condition.

The falsification deliberately sharpens the adversary. The bound of
Sec.~\ref{sec:headB} uses \texttt{flip\_all} as a worst-case upper bound; to hunt
counterexamples we instead use a single flipped sink bit, the sharpest probe of whether
one wrong bit needs a faulty path to propagate. Under it every candidate is refuted
(Table~\ref{tab:falsify}). The canonical reduction (undirected all-faulty connectivity
from the flip to the cut) survives only $0.071$ of propagating shots ($23{,}934$
counterexamples), and essentially none at commit width two or more; the $q^{W/2}$
subsidy refinement survives a minority; only the vacuous floor (some fault present
somewhere in the band) survives (Table~\ref{tab:falsify}). Propagation routinely occurs
with no faulty path anywhere near the flipped bit. The faulty-path route to formalizing
the reduction is therefore closed. (Under \texttt{flip\_all} the same reduction survives
$0.52$, which is why an aggregate certificate alone does not expose the gap.)

% Generated by scripts/make_falsify_table.py from results/falsify.json.
\begin{table*}[t]
\centering
\caption{The modeling hypothesis is false shot by shot. Prop-weighted conditional survival $P(\text{predicate}\mid\text{propagation})$ of each candidate necessary-structure predicate under the sharp single-bit adversary, over the full $d,p,W$ envelope ($25{,}765$ propagating shots). A necessary condition survives at $1$; every structural predicate, including both literal forms of the Lean hypothesis (starred), falls far below it, and only the vacuous floor survives. The two Lean forms collapse further at commit width two or more (\texttt{reach\_all} to $0.005$, \texttt{subsidy\_half} to $0.35$). The weaker a predicate, the better it survives, the signature of a hypothesis with no true faulty-path content.}
\label{tab:falsify}
\begin{tabular}{llr}
\toprule
Predicate & Structural content & Survival \\
\midrule
\texttt{flow2\_faulty} & $\ge 2$ disjoint faulty sink$\to$cut paths & $0.000$ \\
\texttt{reach\_all\_oriented} & monotone all-faulty sink$\to$cut path & $0.068$ \\
\texttt{reach\_all}$^\star$ & undirected all-faulty connectivity (canonical) & $0.071$ \\
\texttt{subsidy\_half}$^\star$ & length-$W$ path with $\ge\lceil W/2\rceil$ faulty edges & $0.256$ \\
\texttt{weighted\_subsidy} & subsidised crossing $\le$ boundary-absorb cost & $0.488$ \\
\texttt{layer\_cover} & faulty cluster crosses every layer (Peierls) & $0.071$ \\
\texttt{fault\_near\_flip\_r2} & faulty edge within graph distance 2 of flip & $0.501$ \\
\texttt{fault\_near\_flip\_r0} & flip incident to a faulty edge & $0.060$ \\
\texttt{any\_faulty} & any faulty band edge (vacuous floor) & $1.000$ \\
\bottomrule
\end{tabular}
\end{table*}

A re-pairing instrument then demonstrates the real mechanism (Table~\ref{tab:mech}).
Decoding twice (baseline and single-bit flip) and taking the symmetric difference of
the matched edge sets gives the change in the correction. On propagating shots that
change is a long re-route that reaches the cut depth ($98.8\%$ of the time) carrying no
faulty flip-to-cut path ($99.5\%$ of the time), while on non-propagating shots it is a
single edge, a local absorb at the boundary. The crossing is cheaper than the absorb:
the re-route re-pairs an existing cut-region defect with the flipped bit, recouping that
defect's matching cost. Propagation is a global minimum-weight re-pairing, not
faulty-path percolation. The per-rate breakdown matters and the pooled medians are
confounded, since $94\%$ of propagating events occur at $p=10^{-2}$: the direction, a
longer re-route at smaller matching weight reaching the cut with no faulty path, holds
at every $p$. The mechanism is partly degeneracy-driven, and increasingly so with $p$:
the fraction of decisions that flip under a small weight perturbation rises from $0.04$
at $p=10^{-3}$ to $0.40$ at $p=10^{-2}$, so at high noise the re-pairing is frequently a
near-tie while at low noise it is a cleaner but far rarer win. The low-$p$ row rests on
only $24$ propagating events and is indicative.

% Generated by scripts/make_mechanism_table.py from results/mechanism.json.
\begin{table*}[t]
\centering
\caption{The real mechanism is a global minimum-weight re-pairing, per noise rate $p$ (single-bit worst-case adversary). On propagating shots the correction re-routes a long path (large symmetric-difference, many edges) at a \emph{smaller} matching-weight change $\Delta w$ than the local absorb on non-propagating shots (prop.\,/\,non-prop.\ medians), reaching the cut depth with no faulty flip-to-cut path. The direction holds at every $p$; the tie-sensitive fraction (decisions that flip under a small weight perturbation) rises with $p$, so at high noise the re-pairing is frequently a near-tie (degeneracy-driven) while at low noise it is a cleaner but far rarer win. Pooled over all $p$ the re-route reaches the cut on $0.988$ of propagating shots with a faulty path absent on $0.995$; the low-$p$ row rests on few events and is indicative.}
\label{tab:mech}
\begin{tabular}{lrccccc}
\toprule
$p$ & $n_{\mathrm{prop}}$ & symdiff (p\,/\,np) & $\Delta w$ (p\,/\,np) & reaches cut & no faulty path & tie-sensitive \\
\midrule
$10^{-3}$ & 24 & 5\,/\,1 & 3.17\,/\,5.05 & $1.000$ & $1.000$ & $0.04$ \\
$3\times10^{-3}$ & 348 & 6\,/\,1 & 2.48\,/\,3.95 & $0.971$ & $0.994$ & $0.13$ \\
$10^{-2}$ & 5771 & 11\,/\,1 & 1.41\,/\,2.44 & $0.989$ & $0.995$ & $0.40$ \\
\bottomrule
\end{tabular}
\end{table*}

None of this touches the Lean core, which is a correct conditional theorem. What is
falsified is the hypothesis it is conditional on, so the exponential containment is real
but is not explained by faulty-path percolation. The correct future bound is a
statement about matching weight: propagation needs $\Omega(W)$ subsidy to beat the
local absorb over $W$ rounds, which is plausibly the source of the measured
$P_{\mathrm{prop}}\sim\lambda^{W}$ but is not proved here (Sec.~\ref{sec:limits}).

\section{Decide: the speculation-decision pass}
\label{sec:headC}

To make the physical blast-radius bounds and predictability limits actionable for a runtime software stack, we abstract them into a compiler pass. The two numbers above feed a compiler decision. We work at the level the decision
lives, an abstract window dependency graph, with windows on a space-by-time grid and
dependencies typed temporal or spatial. This graph makes no assumption about physical
qubit topology or proprietary control hardware: it captures only the window-to-window
commit dependencies, so the pass is portable across code layouts and control stacks. A
front end that lowers a routed program to this graph is a thin future adapter and is not
needed to make the decision. A transpiler
pass annotates each boundary with a speculate flag from the cost model
\begin{equation}
\mathrm{cost} = \mathrm{predictability}\times\mathrm{slack} - \mathrm{blast\_radius},
\end{equation}
wiring the real head-A predictability ($0.992$ at $d=13$) and the head-B couplings
(temporal $1.46\times10^{-7}$, spatial $0.064$), where the blast-radius term is the
expected recovery cost and the restart set is itself derived from the couplings.

The pass derives SWIPER's policy rather than hardcoding it: speculate on all
boundaries, and on a misprediction restart the poisoned window (temporal radius zero)
plus its corner-adjacent neighbor (spatial radius one). This holds for any restart
threshold between the two couplings, a window of $4.4\times10^{5}$, because the
blast-radius work separated the temporal and spatial couplings by more than six orders
of magnitude; a counterfactual that raises the temporal coupling to $0.1$ expands the
temporal restart radius to one, so the policy follows the numbers. On a mixed graph with
a low-predictability region the cost model beats both speculate-everything and
speculate-nothing.

\section{Execute: the runtime executor}
\label{sec:headD}

The decision is only a proxy until something executes it. A standalone executor takes
the pass's plan and runs predict, verify, and bounded-recover against the real harness,
turning the cost-model proxy into a measurement. The latency model is stated and
hardware-agnostic: each window decode costs $L$, a non-speculated boundary forces its
successor to wait (cost $L$ on the critical path), and a speculated boundary starts
immediately except for the expected restart, which carries the measured misprediction
rate times the measured propagation rate. Makespan is the longest path over the window
graph.

Recovery is exact and bounded (Table~\ref{tab:exec}). Under a worst-case misprediction
the committed decode is correct on every shot, over $40{,}000$ shots per cell, because
the lazy verify is the full decode and the bounded restart restores it. The recover
loop converges in a single round: a radius-zero restart of the poisoned window suffices.
The loop is bounded in its header, so termination is structural.

The measurement that matters is the restart penalty, since the speedup over no
speculation on a chain of $N$ windows is $N$ up to that penalty by construction. The
penalty is small: the per-boundary expected restart cost is the measured misprediction
rate times the measured propagation rate, about $1.3\times10^{-5}$, because
mispredictions are rare and almost all are absorbed. On a chain of $16$ windows this
gives a speedup of $15.99$ to $16.00$ across distances and commit widths, holding even
when predictability is degraded to $0.7$ (Table~\ref{tab:exec}), so the policy removes
the per-window commit-chain stall up to a penalty of order $10^{-5}$. The ideal for an
$N=16$ chain is exactly $16$, every window's commit-chain wait hidden, so the executor
runs at $99.9\%$ of the architecture's theoretical limit, the residual being the restart
penalty. This is the critical-path dependency latency, not a wall-clock runtime claim,
which depends on the commit-chain fraction of total runtime (SWIPER's regime).

% Generated by scripts/make_executor_table.py from results/executor.json.
\begin{table*}[t]
\centering
\caption{Executing the policy on the real harness ($N=16$ window chain, $p=10^{-3}$, over $40{,}000$ shots per cell). The speedup over no speculation sits at the ideal chain length ($16$) up to a restart penalty of order $10^{-5}$, holding even when predictability is stressed to $0.7$; the restart-trigger rate (a misprediction that propagates) falls with the commit width $W$; and the committed decode is correct on every shot. The counterfactual restart-cost sweep is discussed in the text.}
\label{tab:exec}
\begin{tabular}{ccrccc}
\toprule
$d$ & $W$ & Restart rate & Speedup & Speedup (pred $=0.7$) & Recovery \\
\midrule
 7 & 1 & $8.9\times10^{-3}$ & $15.999$ & $15.80$ & exact \\
 7 & 2 & $1.3\times10^{-3}$ & $16.000$ & $15.94$ & exact \\
 7 & 3 & $1.0\times10^{-4}$ & $16.000$ & $15.99$ & exact \\
 7 & 4 & $0$ & $16.000$ & $15.99$ & exact \\
 9 & 1 & $1.3\times10^{-2}$ & $15.997$ & $15.81$ & exact \\
 9 & 2 & $2.4\times10^{-3}$ & $15.999$ & $15.93$ & exact \\
 9 & 3 & $3.2\times10^{-4}$ & $16.000$ & $15.99$ & exact \\
 9 & 4 & $1.5\times10^{-4}$ & $16.000$ & $16.00$ & exact \\
\bottomrule
\end{tabular}
\end{table*}

Executing the plan also corrects the cost model. The head-C blast-radius term assumes
every misprediction restarts, but the executor measures that only the propagating
fraction does, so the true expected restart cost is smaller by the propagation
probability, a factor of $10^{-3}$ at small commit width and $10^{-7}$ at $W=d$. At the
real restart cost the corrected and proxy decisions agree and the head-C result stands;
the gap opens only when restarts are made artificially expensive, where the proxy
declines speculation prematurely (a speedup of $1.78\times$ where $14.5\times$ is
available) and the corrected decision stays optimal. Folding the propagation
probability into the cost model is a one-line refinement.

\section{Decoder-relativity: what is decoder-agnostic}
\label{sec:relativity}

The contribution sits above the decoder, so we test which findings depend on the
decoder. We build a second, algorithmically distinct decoder, the union-find decoder of
Delfosse and Nickerson~\cite{unionfind} (unweighted cluster growth and peeling), on the
same matching graph the MWPM reference uses, so only the matching rule changes. It is
validated first: every correction reproduces the syndrome exactly (node-degree parity
equals syndrome on every shot), and the logical error rate suppresses with distance,
tracking MWPM within a small factor. Because this union-find is unweighted its threshold
is below MWPM's, so the cross-decoder check is scoped to $p\le 3\times10^{-3}$, the
regime where it is a validated below-threshold decoder.

\begin{figure}[t]
\includegraphics[width=\columnwidth]{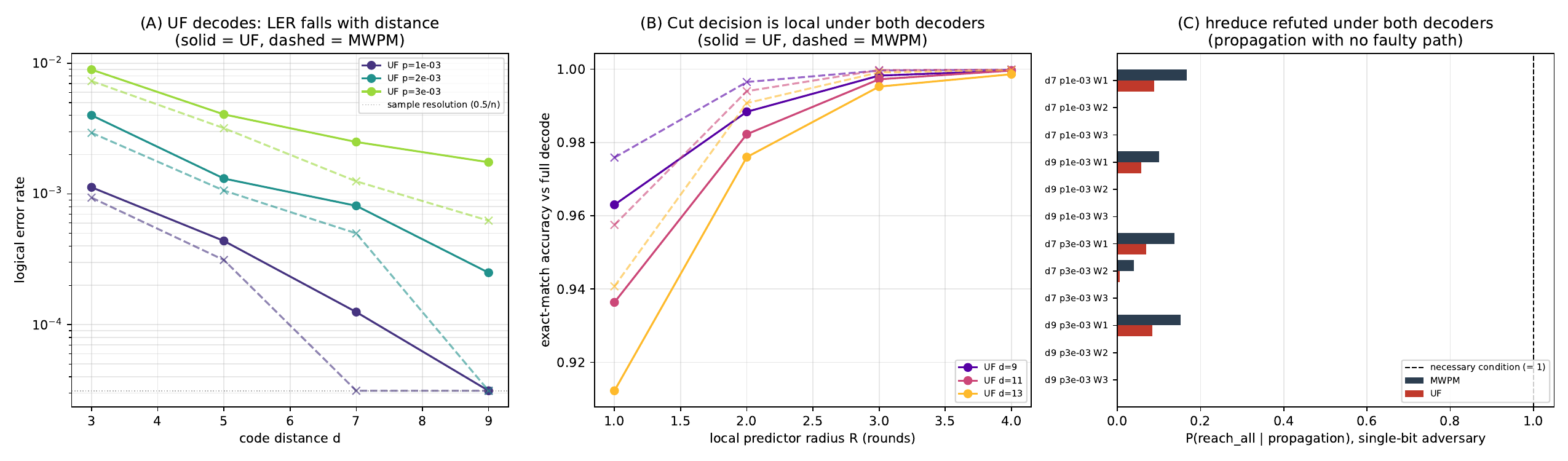}
\caption{Decoder-relativity under a second decoder (union-find). (Left) union-find is a
validated below-threshold decoder: its logical error rate falls with distance. (Center)
the cut decision is local under both decoders, saturating by a receptive field of three
rounds. (Right) propagation with no faulty path persists under both decoders, so the
refutation is decoder-robust.}
\label{fig:relativity}
\end{figure}

Two structural findings are decoder-robust (Fig.~\ref{fig:relativity}). The locality of
the cut decision (Sec.~\ref{sec:headA}) holds, checked here over $d=9$ to $13$:
radius-$R$ accuracy saturates near one by $R=3$ under union-find as under MWPM, about
one to two points noisier with the same shape. The refutation of the reduction
(Sec.~\ref{sec:falsify}) holds: the reduction survives at most $0.005$ of propagating
shots at commit width two or more under union-find, and union-find propagates more
events yet refutes at least as hard.

What is decoder-relative is the absolute predictability magnitude and the speedup
ratios, which depend on the reference decoder's per-bit accuracy, the rate constant
$\lambda$, and the minimum-weight re-pairing mechanism of Sec.~\ref{sec:falsify}, which
is intrinsic to a minimum-weight decoder and so was deliberately not re-run under
union-find. The precise statement is this: the predict/verify/recover wrapper is
decoder-agnostic by construction, since any decoder plugs in behind the verify step and
union-find does so unchanged; the phenomenology is decoder-robust at the level a runtime
needs (locality, no-faulty-path propagation, radius one) and decoder-relative in its
absolute numbers and in the min-weight explanatory mechanism.

\section{Limitations and future work}
\label{sec:limits}

\textit{The matching-weight bound.} The falsification closed the faulty-path route to
formalizing the reduction, and the mechanism study shows the correct statement is about
matching weight: propagation needs $\Omega(W)$ subsidy to beat the local absorb over $W$
rounds. That is the natural source of the measured $P_{\mathrm{prop}}\sim\lambda^{W}$,
but it is not proved; it is the next formalization target and is out of scope here.

\textit{Sim-accurate speedup.} The speedup numbers in Sec.~\ref{sec:headA} are the
analytic linear-chain map with a free speculation depth. The graph-structured,
restart-cost-accurate speedup needs the SWIPER-SIM pipeline, which is not public; that
validation is future work and the natural contribution back to SWIPER-SIM.

\textit{Spatial blast radius.} The corner coupling is characterized (a bounded few
percent, corner-localized, width-invariant) but not bounded the way the temporal
direction is.

\textit{Decoder-relativity envelope.} Union-find is unweighted, so the cross-decoder
check is scoped to $p\le 3\times10^{-3}$. A weighted-growth union-find for higher $p$,
and a non-matching decoder with a redefined boundary readout, are future work. The
min-weight mechanism is matching-specific by construction.

\textit{Harness reconstruction.} SWIPER-SIM's exact configuration is not public, so the
harness is a documented reconstruction and all headroom is internal
(predictor-versus-bracket on the same harness). A direct SWIPER-SIM cross-check is
future work.

\textit{Bare-metal deployment.} The executor here runs the predict-verify-recover loop
in software against the harness, measuring critical-path dependency latency under a
stated hardware-agnostic cost model. Deploying the same wrapper on a physical control
kernel or an FPGA-based real-time decoding environment, where the round cadence and
decode latency are set by hardware rather than modeled, would turn the critical-path
result into a wall-clock one and is the natural next step.

\section{Conclusion}

The predictor half of speculative window decoding was built twice; the verifier half
was open. We built it: a predictability bracket that shows the cross-boundary decision
is local and SWIPER is near the achievable accuracy, a worst-case blast-radius bound
with a machine-checked probability core that shows a wrong guess is contained to radius
one and adds no error floor, a falsification and mechanism study that replaces the
bound's modeling hypothesis with a global minimum-weight re-pairing (demonstrated where
propagation is common, at $p=10^{-2}$, and directionally robust though on few events at
lower noise), a compiler pass that derives SWIPER's restart policy from these numbers,
and a runtime executor that confirms on the real harness that the loop recovers exactly
and removes the commit-chain stall up to a small restart penalty. The verifier side is a
reusable layer that sits above any decoder; a second
decoder confirms that the structural results are decoder-robust while the absolute
magnitudes and the min-weight mechanism are specific to matching. The code, the
machine-checked development, and every figure and table are reproducible from fixed
seeds.

\section*{Code and data availability}

The harness, predictors, blast-radius operator, compiler pass, runtime executor,
union-find decoder, the Lean~4 development, and the scripts that regenerate every result,
figure, and table from fixed seeds are available at
\url{https://code.rylanmalarchick.com/rylanmalarchick/speculative-qec}.

\section*{Acknowledgments}

The path-count bound in the Lean development was closed with the Aristotle automated
prover.

\section*{AI Disclosure}

AI-assisted tools (Claude, Anthropic) were used for code development, running the
numerical experiments, and manuscript preparation. The research direction, design
decisions, and interpretation are the author's, who takes full intellectual
responsibility for all content.

\bibliographystyle{unsrt}
\bibliography{refs}

@inproceedings{swiper,
  author    = {Viszlai, Joshua and Chadwick, Jason D. and Joshi, Sarang and Ravi, Gokul Subramanian and Li, Yanjing and Chong, Frederic T.},
  title     = {{SWIPER}: Minimizing Fault-Tolerant Quantum Program Latency via Speculative Window Decoding},
  booktitle = {Proceedings of the 52nd Annual International Symposium on Computer Architecture (ISCA)},
  year      = {2025},
  doi       = {10.1145/3695053.3731022},
  note      = {arXiv:2412.05115},
}

@inproceedings{artery,
  author    = {Tian, Wuwei and Lu, Liqiang and Tan, Siwei and Liang, Yun and Li, Tingting and Zhou, Kaiwen and Jia, Xinghui and Yin, Jianwei},
  title     = {{ARTERY}: Fast Quantum Feedback using Branch Prediction},
  booktitle = {Proceedings of the 52nd Annual International Symposium on Computer Architecture (ISCA)},
  year      = {2025},
  doi       = {10.1145/3695053.3731086},
}

@article{battistel_realtime,
  author  = {Battistel, Francesco and Chamberland, Christopher and Johar, Kauser and Overwater, Ramon W. J. and Sebastiano, Fabio and Skoric, Luka and Ueno, Yosuke and Usman, Muhammad},
  title   = {Real-Time Decoding for Fault-Tolerant Quantum Computing: Progress, Challenges and Outlook},
  journal = {Nano Futures},
  volume  = {7},
  pages   = {032003},
  year    = {2023},
  doi     = {10.1088/2399-1984/aceba6},
  note    = {arXiv:2303.00054},
}

@article{caune_realtime,
  author  = {Caune, Laura and Skoric, Luka and Blunt, Nick S. and Ruban, Archibald and McDaniel, Jimmy and Valentine, Joseph A. and others},
  title   = {Demonstrating real-time and low-latency quantum error correction with superconducting qubits},
  journal = {arXiv preprint},
  year    = {2024},
  note    = {arXiv:2410.05202},
}

@article{google_belowthreshold,
  author  = {{Google Quantum AI and Collaborators}},
  title   = {Quantum error correction below the surface code threshold},
  journal = {Nature},
  volume  = {638},
  pages   = {920--926},
  year    = {2025},
  doi     = {10.1038/s41586-024-08449-y},
  note    = {arXiv:2408.13687},
}

@article{terhal_review,
  author  = {Terhal, Barbara M.},
  title   = {Quantum error correction for quantum memories},
  journal = {Reviews of Modern Physics},
  volume  = {87},
  pages   = {307--346},
  year    = {2015},
  doi     = {10.1103/RevModPhys.87.307},
  note    = {arXiv:1302.3428},
}

@article{pinball,
  author  = {Knapen, Alexander and Tao, Guanchen and Mack, Jacob and others},
  title   = {Pinball: A Cryogenic Predecoder for Surface Code Decoding Under Circuit-Level Noise},
  journal = {arXiv preprint},
  year    = {2025},
  note    = {arXiv:2512.09807},
}

@article{parallelwindow,
  author  = {Skoric, Luka and Browne, Dan E. and Barnes, Kenton M. and Gillespie, Neil I. and Campbell, Earl T.},
  title   = {Parallel window decoding enables scalable fault tolerant quantum computation},
  journal = {Nature Communications},
  volume  = {14},
  pages   = {7040},
  year    = {2023},
  doi     = {10.1038/s41467-023-42482-1},
}

@inproceedings{promatch,
  author    = {Alavisamani, Narges and Vittal, Suhas and Qureshi, Moinuddin},
  title     = {{Promatch}: Extending the Reach of Real-Time Quantum Error Correction with Adaptive Predecoding},
  booktitle = {Proceedings of the 29th ACM International Conference on Architectural Support for Programming Languages and Operating Systems (ASPLOS)},
  year      = {2024},
  note      = {arXiv:2404.03136},
}

@inproceedings{leviathan,
  author    = {Leviathan, Yaniv and Kalman, Matan and Matias, Yossi},
  title     = {Fast Inference from Transformers via Speculative Decoding},
  booktitle = {Proceedings of the 40th International Conference on Machine Learning (ICML)},
  year      = {2023},
  note      = {arXiv:2211.17192},
}

@article{spec_sampling,
  author  = {Chen, Charlie and Borgeaud, Sebastian and Irving, Geoffrey and Lespiau, Jean-Baptiste and Sifre, Laurent and Jumper, John},
  title   = {Accelerating Large Language Model Decoding with Speculative Sampling},
  journal = {arXiv preprint},
  year    = {2023},
  note    = {arXiv:2302.01318},
}

@article{unionfind,
  author  = {Delfosse, Nicolas and Nickerson, Naomi H.},
  title   = {Almost-linear time decoding algorithm for topological codes},
  journal = {Quantum},
  volume  = {5},
  pages   = {595},
  year    = {2021},
  doi     = {10.22331/q-2021-12-02-595},
  note    = {arXiv:1709.06218},
}

@article{stim,
  author  = {Gidney, Craig},
  title   = {Stim: a fast stabilizer circuit simulator},
  journal = {Quantum},
  volume  = {5},
  pages   = {497},
  year    = {2021},
  doi     = {10.22331/q-2021-07-06-497},
  note    = {arXiv:2103.02202},
}

@article{pymatching,
  author  = {Higgott, Oscar and Gidney, Craig},
  title   = {Sparse Blossom: correcting a million errors per core second with minimum-weight matching},
  journal = {Quantum},
  volume  = {9},
  pages   = {1600},
  year    = {2025},
  doi     = {10.22331/q-2025-01-20-1600},
  note    = {arXiv:2303.15933},
}

@article{topological_memory,
  author  = {Dennis, Eric and Kitaev, Alexei and Landahl, Andrew and Preskill, John},
  title   = {Topological quantum memory},
  journal = {Journal of Mathematical Physics},
  volume  = {43},
  number  = {9},
  pages   = {4452--4505},
  year    = {2002},
  note    = {arXiv:quant-ph/0110143},
}

@article{fowler_surface,
  author  = {Fowler, Austin G. and Mariantoni, Matteo and Martinis, John M. and Cleland, Andrew N.},
  title   = {Surface codes: Towards practical large-scale quantum computation},
  journal = {Physical Review A},
  volume  = {86},
  pages   = {032324},
  year    = {2012},
  doi     = {10.1103/PhysRevA.86.032324},
  note    = {arXiv:1208.0928},
}

@misc{mathlib,
  author       = {{The Mathlib Community}},
  title        = {The {Lean} Mathematical Library},
  howpublished = {Proc. 9th ACM SIGPLAN Int. Conf. on Certified Programs and Proofs (CPP)},
  year         = {2020},
  doi          = {10.1145/3372885.3373824},
}

\end{document}